\documentclass[11pt]{article}
\usepackage{array}
\usepackage{graphicx}
\usepackage{cite}
\usepackage{textpos}
\usepackage{bbold}
\usepackage{adjustbox}

\title{Ambiguity, Invisibility, and Negativity
\footnote{Contribution to  Stanley Deser Memorial Volume ``Gravity, Strings and Beyond"}
\author {Frank Wilczek  \\
\small\it Center for Theoretical Physics, MIT, Cambridge, MA 02139 USA; \\
\small\it T. D. Lee Institute and Wilczek Qtauantum Center, \\
\small\it Shanghai Jiao Tong University, Shanghai, China;\\
\small\it Arizona State University, Tempe, AZ, USA; \\
\small\it Stockholm University, Stockholm, Sweden }}

\begin{document}

\maketitle

\begin{textblock*}{5cm}(11cm,-8.2cm)
\fbox{\footnotesize MIT-CTP/5609}
\end{textblock*}

\begin{abstract}
Many widely different problems have a common mathematical structure wherein limited knowledge lead to ambiguity that can be captured conveniently using a concept of invisibility that requires the introduction of negative values for quantities that are inherently positive.  Here I analyze three examples taken from perception theory, rigid body mechanics, and quantum measurement.
\end{abstract}

\medskip

Stanley Deser's generosity and humor lifted my spirits on many occasions over many years.  Our professional work in physics had very different centers, but there was some overlap.  We even wrote a short paper together \cite{deser_wilczek}.  That paper is a minor work by any standard, though it does touch on a significant point.   In it, we gave several examples of nonabelian gauge potentials that generate the same gauge fields but different gauge structures, so that (for instance) $F^1_{\alpha \beta} = F^2_{\alpha \beta}$ but $\nabla_\gamma  F^1_{\alpha \beta} \neq \nabla_\gamma  F^2_{\alpha \beta}$ . This contrasts with the abelian case, where the fields determine the gauge potentials up to a gauge transformation, locally.  (Globally, of course, they do not \cite{dirac, aharonov_bohm}.)   The problem of classifying the ambiguity in cases like this has no general solution; indeed, the closely related problem of classifying spaces with equal curvature data of different kinds in different dimensions up to isometry quickly points us to some milestone theorems, famous unsolved problems, and unexplored territory.

Here I will describe a trio of more down-to-earth problems that have the same flavor, but which share a mathematical structure that is much more tractable.  In the context of ``Gravity, Strings, and Beyond'' they fall firmly within ``Beyond'',  not in the sense of ``Transcending'', but rather just ``Outside''.   They are sufficiently direct and simple that further introduction seems unwarranted.

\section{Metamers in Visual Perception}

Within the vast and complex subject of visual perception \cite{wandell} there is a useful idealization, with roots in the work of Maxwell \cite{maxwell}, that captures important aspects of the primary perception of color.  This  is called colorimetry.   The book by Koenderink \cite{koenderink} is a very attractive presentation of many aspects of  theoretical colorimetry.  

The central concept of colorimetry is that the primary perception of the color of an illumination source -- essentially meaning, in this context, a uniform beam of light -- can be predicted using a few linear functions of its spectrum.   Thus we summarize the responses of several detectors' $\alpha$ with response functions $c_\alpha (\lambda)$ to different illumination sources $k$ with intensity spectra $I_k(\lambda)$ according to 
\begin{equation}
M_{\alpha k } ~=~ \int \, d\lambda \, c_\alpha (\lambda) I_k (\lambda)
\end{equation}
What we mean by ``predicting'' the primary perception is that illumination sources that induce the same values of $M_{\alpha k}$ will be indistinguishable to the detectors.  This is the possibility we will be analyzing..   

``Normal'' -- i.e., majority -- human color perception is trichromatic.  That is to say, most people share three very similar sensitivity functions, often called ``blue, green, red'' after the location of their peak values.  They are rather broadly tuned, however, and in the scientific literature ``S, M, L'' (for ``short, medium, long'') is generally preferred.  Maxwell did ingenious psychophysical experiments to establish the linearity and three-dimensionality of normal human color perception.   Nowadays we can trace its molecular origin.  There are three basic pigments, concentrated in three types of cone cells in the fovea, that can undergo shape changes upon absorbing photons.  The shape changes trigger electrical impulses that are the primary events in color vision.  These absorption events are probabilisitic and all-or-none.  Human color vision is a beautiful case study in quantum mechanics at work!    

An illumination $I(b_k, \lambda) \equiv \sum\limits_k b_k I_k (\lambda)$ that satisfies
\begin{equation}\label{invisible_metamer_space}
0 ~=~ \int \, d\lambda c_\alpha (\lambda) \, I(b_k, \lambda) ~=~ \sum\limits_k \, M_{\alpha k} b_k 
\end{equation}
will be invisible to all the detectors.  Given $M_{\alpha k}$, conditions (\ref{invisible_metamer_space}) are a system of linear equations for the $b_k$.  Their solutions define a linear space that we will refer to as the space of {\it invisible metamers}.  (The term ``black metamers'' is often used, but -- like ``dark matter'' and ``dark energy'' - it tends to evoke misleading imagery.)

Since the $c_\alpha(\lambda)$ and $I_k(\lambda)$ are intrinsically positive, so are the $M_{\alpha k}$.  To obey Eqn.\,(\ref{invisible_metamer_space}), therefore, some of the $b_k$ will have to be negative.  Since $b_k$ represents the strength with which illumination source $k$ is present, however, only $b_k \geq 0$ are physically realizable.  Nevertheless, the invisible metamer concept is quite useful, because it parameterizes the ambiguity left open by perception.   The point is that two illumination choices $b^{(1)}_k,  b^{(2)}_k$ look the same to all the detectors if and only if $b^{(1)}_k - b^{(2)}_k$ belongs to the space of invisible metamers.  Thus, given any physical illumination choice $b^{\rm phys.}_k$, we can find all the perceptually equivalent by illuminations by adding in vectors from the space of invisible metamers, as $b^{\rm phys.}_k + b^{\rm inv.}_k$.  

The situation becomes richer, and our conceptual clarity bears fruit, when we come to compare different sets of detectors \cite{metamer_paper}.   Let me describe a sample application from that paper.   There are common forms of variant color perception, usually called color blindness, that result from mutations of the S, M, or L receptor molecules.   Now suppose that we want to make a differential diagnosis among them.  The invisible metamer concept suggests a powerful and efficient way to do that.   Indeed, if we have four illumination sources (say four types of LEDs) with adjustable brightness, then there will be {\it different\/} one-dimensional invisible metamer spaces associated with the normal and variant receptor sets.  Let us call the basis vectors $b^{N}_k, b^{S^\prime}_k, b^{M^\prime}_k, b^{L^\prime}_k$, in an obvious notation.  Then, starting with a reference color combination $b^{\rm O}_k$ that has all positive components, we can mix dial in illumination patterns of the types
\begin{eqnarray}
{\rm normal \ metamers}: &{}& b^{\rm O} + \lambda b^{N} \nonumber \\
{\rm S \ mutant \ metamers}: &{}& b^{\rm O} + \lambda b^{S^\prime} \nonumber \\
{\rm M \ mutant \ metamers}: &{}& b^{\rm O} + \lambda b^{M^\prime} \nonumber \\
{\rm L \ mutant \ metamers}: &{}& b^{\rm O} + \lambda b^{L^\prime}
\end{eqnarray}
with variable $\lambda$.  The first type will provide, for different values of $\lambda$, a set of colors that cannot be distinguished by normal trichromats, but that {\it are\/} distinguishable by the mutants.   This phenomenon shows, rather dramatically, why it is not entirely appropriate to refer to the mutations as ``color blindness''.   The second type provides colors that cannot be distinguished by S mutants, but can be distinguished by normal trichromats and M or L mutants, and so forth.  By choosing appropriate illumination sources we can accentuate the differences.  Following this strategy, we have made good, simple practical devices.  

Along similar lines, one can design quantitative tests for different hypothetical forms of ``super'' color vision.  Indeed, since the relevant genes lie on the X chromosomes, females (with two X chromosomes) can carry both majority and mutant forms of the different receptors, allowing different kinds of tetrachromacy or even pentachromacy.   For more on this and other applications, see \cite{metamer_paper}.

\section{Equivalent Rigid Bodies}

In classical mechanics, a rigid body is defined by a distribution of masses $m_j$  in space, at positions $x_j^\alpha$,.   According to the definition of a rigid body, we only consider motions that correspond to common rotation and translation of all the masses, induced by given summed forces (and torques).   The degrees of freedom can be taken as the overall position and orientation of a ``body-fixed'' reference system.  

As is shown in textbooks, the dynamics of a rigid body -- i.e., the evolution of its position and orientation --  depends only on its total mass and its inertia tensor
\begin{equation}
I^{\alpha \beta} ~=~ \sum\limits_j \, m_j ( | x_j |^2 \delta^{\alpha \beta} - x_j^\alpha x_j^\beta) 
\end{equation}
referred to a coordinate system where the center of mass
\begin{equation}
x^\alpha_{\rm CM} ~=~ \frac{\sum\limits_j m_j x_j^\alpha}{ \sum\limits_j m_j} 
\end{equation}
is at the origin.
It is possible for different distributions of mass, i.e. different bodies, to agree in those properties.  In that case, if we have access only to those bodies' overall motion - for example, if they are rigidly attached within identical opaque shells - then we will not be able to distinguish them.  We can say that they are dynamically equivalent.  

The problem arises, to clarify and exemplify this ambiguity mathematically.

The conditions for equality of total mass and inertia tensors, and zeroing of centers of mass, are all linear in the component mass variables $m^j$.  It is therefore natural, by analogy to our treatment of metamerism, to introduce a space of ``dynamically invisible bodies''.  Dynamically invisible bodies are defined by distributions of mass  such that
\begin{eqnarray}\label{invisible_bodies}
0 ~&=&~ \sum\limits_j m_j  \nonumber \\
0 ~&=&~ \sum\limits_j m_j x_j^\alpha \nonumber \\
0 ~&=&~ \sum\limits_j \, m_j ( | x_j |^2 \delta^{\alpha \beta} - x_j^\alpha x_j^\beta) 
\end{eqnarray}

In order for Eqn.\,(\ref{invisible_bodies}) to be satisfied some of the $m_j$ will need to be negative.  Thus, dynamically invisible bodies, like invisible metamers, are not directly physical.  But dynamically invisible bodies are relatively simple to construct, because their defining conditions are linear and highly symmetric.  Dynamically invisible bodies are a useful conceptual tool, because we can construct physical dynamically equivalent objects from dynamically invisible bodies  (i.e., their mass distributions) by adding invisible bodies to a positive mass distribution.  

Simple but flexible constructions based on these ideas can be used to generate complex, non-obvious examples of dynamically equivalent bodies.  Here are two such constructions:
\begin{enumerate}
\item {\it Parity construction}: To any distribution of masses $m_j$ at positions $x_j^\alpha$,  $j = 1, ..., n$ whose center of mass is at the origin, add reflected negative masses at the inverted positions, according to
\begin{eqnarray}
m_{-j} ~&=&~ - m_j \nonumber \\
x_{-j}^\alpha ~&=&~ - x_j^\alpha 
\end{eqnarray} 
This creates a dynamically invisible body.
\item {\it Rotation construction}: To any distribution of masses $m_j$ at positions $x_j^\alpha$,  $j = 1, ..., n$ whose center of mass is at the origin, and whose inertia tensor is proportional to the unit tensor, and any rotation $R^\alpha_\beta$, add negative masses at the rotated positions, according to 
\begin{eqnarray}
m_{-j} ~&=&~ - m_j \nonumber \\
x_{-j}^\alpha ~&=&~ R^\alpha_\beta x_j^\beta 
\end{eqnarray} 
Here we can allow improper rotations, or use an equal-mass, equal-inertia tensor body of different form. 

Naturally, this begs the question of constructing non-trivial distributions whose inertia tensor is proportional to the unit tensor.  Mass distributions that are symmetric under appropriate discrete subgroups of the rotation group, such as the symmetry groups of the Platonic solids, will have that property.   
\item{\it Superposition}: The invisible bodies form a linear manifold: their mass distributions can be multiplied by constants, and added together freely.
\end{enumerate}

To ground the discussion, let us consider a minimal example of an invisible body. We put masses $m_1 \equiv m, m_2 = ml_1/l_2$ at positions $l_1 \hat z, -l_2 \hat z$.  The parity construction gives us a dynamically invisible body if we add in $m_3 = -m, m_4 = - ml_1/l_2$  at positions $- l_1 \hat z, l_2 \hat z$. Now if we add this to a mass distribution
$M_1, M_2, M_3, M_4$ at $l_1 \hat z, -l_2 \hat z, - l_1 \hat z, l_2 \hat z$ and
\begin{eqnarray}
M_3 ~&\geq&~ m \nonumber \\
M_4 ~&\geq&~ m l_1/l_2
\end{eqnarray}
we will define a physical mass distribution.  By varying $m>0$ within these constraints, we produce a family of dynamically equivalent physical mass distributions.

Untethered point masses are an extreme idealization of any actual rigid body, of course.  We can make the foregoing construction more realistic by replacing the point masses with distributions of mass around the same centers, and by adding supporting material whose mass distribution is independent of $m$ to fill the interstices.   In this way, we reach practically realizable designs for dynamically equivalent rigid bodies.

\bigskip

\section{Quantum Grey Boxes}

The state of a system in quantum mechanics is specified by a density matrix $\rho$, which is required to be Hermitian and non-negative, with unit trace.  Observables are represented by hermitian operators $M$, and the expectation value of $M$ in the state described by $\rho$ is ${\rm Tr}\, \rho M$.   Thus when a suite of measurements of the observables $M_j$ on a system yield results $v_j$, we learn
\begin{equation}\label{measurement_results}
{\rm Tr\/} \rho M_j ~=~ v_j
\end{equation}
These results might not determine $\rho$ completely, and the issues arises, to parameterize the resulting ambiguity.  (The measurements take us from a black box to a grey box.)

Clearly, there is a strong family resemblance among this problem, the preceding one, and the color metamer problem.  Following the same line of thought, we define a linear space of invisible density matrices consisting of hermitian matrices $\rho^{\rm inv.}$ that obey the equations
\begin{eqnarray}
{\rm Tr\/} \rho^{\rm inv.} ~&=&~ 0 \nonumber \\
{\rm Tr\/} \rho^{\rm inv.} M_j ~&=&~ 0
\end{eqnarray}
Invisible density matrices cannot be non-negative, so they do not describe physically realizable states.  Basically, they contain negative probabilities.

An extremely simple example may be helpful here, to ground the discussion.  For a two-level system physical density matrices have the form
\begin{equation}
\rho ~=~ \left(\begin{array}{cc}a & \beta \\ \beta^* & 1-a\end{array}\right)
\end{equation}
where $0 \leq a \leq 1$ is a real number and $\beta$ is a complex number, subject to the constraint
\begin{equation}
a(1-a) - |\beta|^2 \geq 0
\end{equation}
For measurement of $\sigma_3$, the invisible state conditions for the hermitian matrix $M \equiv \left(\begin{array}{cc}r &  \gamma \\ \gamma^* & s \end{array}\right)$ read
\begin{eqnarray}
{\rm Tr\/} M ~&=&~ r + s ~=~ 0 \nonumber \\
{\rm Tr\/} \left(\begin{array}{cc}1 & 0 \\0 & -1 \end{array}\right) M ~&=&~ r-s ~=~ 0
\end{eqnarray}
so 
\begin{equation}
M ~=~ \left(\begin{array}{cc}0 & \gamma \\ \gamma^* & 0\end{array}\right) ~=~ {\rm Re}\, \gamma \ \sigma_1 - {\rm Im}\,  \gamma \ \sigma_2
\end{equation}
Thus, we see that the space of invisible density matrices a spanned by a mixture of spin up and spin down in the $\hat x$ direction with equal and opposite probabilities, together with a mixture of spin up and spin down in the $\hat y$ direction, with equal and opposite probabilities.

Suppose that we measure the expectation value of $\sigma_3$ in the state represented by $\rho$ to be $v$, i.e.
\begin{equation}
{\rm Tr\/} \,  \left(\begin{array}{cc}1 & 0 \\0 & -1 \end{array}\right)  \left(\begin{array}{cc}a & \beta \\ \beta^* & 1-a\end{array}\right) ~=~ 2a -1 ~=~ v
\end{equation}
This leaves $\beta$ undetermined.  Evidently, that ambiguity corresponds to motion within the space of invisible states.  But (noting $a = \frac{v+1}{2}$) the physical states must obey
\begin{equation}
1 - v^2  \geq 4 | \beta|^2
\end{equation}
Thus we can only make use of a portion of the invisible state space, whose extent depends on $v$.

Negative probabilities as they appear in different contexts were the subject of a very entertaining presentation by Feynman, written up in \cite{feynman}. Here, they can offer the same sorts of mathematical convenience and conceptual clarity here as do the invisible metamers and invisible bodies in their contexts; and we can take over ideas from one problem to the others.  We can construct distinct physically realizable density matrices that cannot be resolved by a given measurement suite, or we can compare the blind spots of different measurement suites, for example.  A natural extension will bring in superdensity matrices \cite{cotler_fw} and time-dependent measurements.   Then we will have a precise concept of invisible histories, which arise in any realistic measurement protocol.

\bigskip
{\it Acknowledgements}: Thanks to Nathan Newman and Jordan Cotler for helpful comments.  This work is supported by the U.S. Department of Energy under grant Contract  Number DE-SC0012567, by the European 
Research Council under grant 742104, and by the Swedish Research Council under Contract No. 335-2014-7424.

\end{document}